\documentclass[aps,twocolumn,showpacs]{revtex4}
\usepackage{graphicx}

\begin{document}

\title{Spatial infinity in higher dimensional spacetimes}

\author{Tetsuya Shiromizu$^{(1,2)}$ and Shinya Tomizawa$^{(1)}$}

\affiliation{$^{(1)}$Department of Physics, Tokyo Institute of Technology, 
Tokyo 152-8551, Japan}

\affiliation{$^{(2)}$Advanced Research Institute for Science and Engineering, 
Waseda University, Tokyo 169-8555, Japan}

\date{\today}

\begin{abstract}
Motivated by recent studies on the uniqueness or non-uniqueness of higher dimensional black hole spacetime, 
we investigate the asymptotic structure of spatial infinity in $n$-dimensional spacetimes($n \geq 4$). 
It turns out that the geometry of spatial infinity does not have maximal symmetry due to the non-trivial 
Weyl tensor ${}^{(n-1)}C_{abcd}$ in general. 
We also address static spacetime and its multipole moments $P_{a_1 a_2 \cdots a_s}$. 
Contrasting with four dimensions, we stress that the local structure of spacetimes cannot be unique under fixed a multipole 
moments in static vacuum spacetimes. For example, we will consider the generalized Schwarzschild spacetimes which are deformed 
black hole spacetimes with the same multipole moments as spherical Schwarzschild black holes. To specify 
the local structure of static vacuum solution 
we need some additional information, at least, the Weyl tensor ${}^{(n-2)}C_{abcd}$ at spatial infinity. 
\end{abstract}


\maketitle

\label{sec:intro}
\section{Introduction}

The fundamental study of higher dimensional black holes is gaining importance due to 
TeV gravity \cite{ADD,LHC} and superstring theory. In four dimensions, the no-hair theorem \cite{Price} 
and uniqueness theorems \cite{Israel} are the main results obtained during the golden age 
of study of black hole physics. 
Here we have a question about black holes in higher dimensions.  What about the uniqueness theorem? 
Recently a static black hole has been proven to be unique in higher dimensional and asymptotically 
flat spacetimes \cite{GIS,Next,Old}. However, we cannot show the uniqueness of stationary black holes. This is 
because there is a counter example, that is, there are higher dimensional Kerr solutions \cite{BH} and 
black ring solutions \cite{Reall} which have the same mass and angular momentum parameters. See also 
Ref. \cite{Reall2} for a related issue of supersymmetric black holes. Even if we 
concentrate on static spacetimes, the asymptotic boundary conditions are not unique \cite{GIS}. 
Indeed, we could have a generalized Schwarzschild solution which is not asymptotically 
flat. There are also important 
issues about the final fate of the unstable black string  or stable configuration of Kaluza-Klein black 
holes \cite{GL}. They are still under invetigation. 

In this paper we focus on the fundamental issue of the asymptotic structure of spatial infinity, 
which is closely related to the asymptotic boundary condition in the uniqueness theorem and 
numerical study. See Ref. \cite{Scri} 
for null infinity in higher dimensions, but with a different motivation. First we 
investigate the geometrical structure of spatial infinity in higher dimensions. Spatial 
infinity is essentially an $(n-1)$-dimensional manifold in general $n$-dimensional spacetimes. In four dimensions, 
it should be restricted to being a three dimensional unit timelike hyperboloid with maximal symmetry \cite{AR}. 
In higher dimensions, as shown later, there are many varieties due to the non-trivial 
$(n-1)$-dimensional Weyl tensor. Next, we discuss the higher multipole moments in static spacetimes. 
For four dimensional spacetimes, the local structure of static and vacuum spacetime is uniquely determined by specifying all  
the multipole moments \cite{Beig}. On the other hand, as we see later, 
higher dimensional static spacetimes cannot be fixed by 
multipole moments alone. We need some additional information to fix the spacetimes. One of them is 
the $(n-2)$-dimensional Weyl tensor on the surface normal to the radial direction. 

The rest of this paper is organized as follows. In Sec. II, we define the spatial infinity 
following Ashtekar and Romano \cite{AR}, and then discuss the leading structure of spatial infinity. 
In Sec. III, we concentrate on static spacetimes and again define spatial infinity 
on spacelike hypersurfaces. Then we define and discuss the multipole moments following Geroch \cite{Geroch}. 
Finally, we give a discussion and summary in Sec. IV. 

\label{sec:ATI}

\section{Structure of spatial infinity}

\subsection{Definition}

We begin with the definition of spatial infinity by Ashtekar and Romano \cite{AR}. 
If one is interested only in spatial infinity, their definition is useful. 

{\it Definition.} 
Physical spacetime $(\hat M, \hat g_{ab})$ has a spatial infinity $i_0$ if there is a 
smooth function $\Omega$ satisfying the following features (i) and (ii) and the 
energy-momentum tensor satisfies the fall off condition (iii).

(i) $\Omega \hat =0$ and $ d \Omega| \hat{\neq} 0$ 

(ii)The following quantities have smooth limit on $i_0$:
%
\begin{eqnarray}
 q_{ab}=\Omega^2 (\hat g_{ab}-\Omega^{-4}F^{-1}\hat \nabla_a \Omega \hat \nabla_b \Omega)
=\Omega^2 \hat q_{ab} 
\end{eqnarray}
%
%
\begin{eqnarray}
n^a: = \Omega^{-4} \hat g^{ab} \hat \nabla_b \Omega,
\end{eqnarray}
%
where
%
\begin{eqnarray}
F= \Omega^{-4}\hat g^{ab} \hat \nabla_a \Omega \hat \nabla_b \Omega = \mbox \pounds_n \Omega. 
\end{eqnarray}
%
and $\hat =$ denotes evaluation on $i_0$. $q_{ab}$ has the signature $(-,+,+,\cdots,+)$. 

(iii)$\hat T_{\mu\nu}:= \hat T_{ab} e_\mu^a e_\nu^b =O(\Omega^{2+m})$ near $i_0$, where 
$\lbrace \hat e_\mu^a \rbrace_{\mu=0,1,2,\cdots,n-1}$ is a quasi orthogonal basis of the 
metric $\hat g_{ab}$ and $m>0$. 
The definition is exactly the same as that in four dimensions. 

We write the physical metric in terms of the 
quasi orthogonal basis 
%
\begin{eqnarray}
\hat g^{ab}=\hat n^a \hat n^b + \hat e^a_I \hat e^{bI}, 
\end{eqnarray}
%
where 
%
\begin{eqnarray}
\hat n^a=-\frac{n^a}{{\sqrt {g(n, n)}}}= 
-\Omega^2 F^{-\frac{1}{2}}n^a
\end{eqnarray}
%
and
%
\begin{eqnarray}
\hat e^a_I =  e^a_I \Omega 
\end{eqnarray}
%
$e^a_I$ represents the parts of the quasi orthogonal basis of $\hat q_{ab}$.

\subsection{Leading order structure}

From the above the asymptotic 
behavior near $i_0$ is determined by the regular quantities $q_{ab}$ and $n^a$. 
For example, the extrinsic curvature $\hat K_{ab}$ of $\Omega=$constant 
surfaces is written as  
%
\begin{eqnarray}
\hat K_{ab}=\frac{1}{2} \mbox \pounds_{\hat n} \hat q_{ab} 
= \Omega^{-1} F^{\frac{1}{2}} q_{ab}
-\frac{1}{2}F^{-\frac{1}{2}} \mbox \pounds_{n}  q_{ab}.
\end{eqnarray}
%
Since it is not regular at $\Omega=0$, we defined the regular tensor $K_{ab}$ as 
%
\begin{eqnarray}
K_{ab}=:\Omega \hat K_{ab}=  F^{\frac{1}{2}}  q_{ab}-\frac{1}{2}\Omega F^{-\frac{1}{2}} \mbox \pounds_{n} q_{ab}.
\end{eqnarray}
%
Then we see that 
%
\begin{eqnarray}
K_{ab} \hat = F^{\frac{1}{2}}  q_{ab}. \label{K}
\end{eqnarray}
%

In the physical spacetime, the Codacci equation is 
%
\begin{eqnarray}
\hat e^a_I \hat n^b \hat T_{ab}= \Biggl[\hat D_b \hat K^b_a -\hat D_a \hat K  \Biggr]\hat e^a_I. 
\end{eqnarray}
%
It is also expressed as 
%
\begin{eqnarray}
\Omega^{-2} \hat e^a_I \hat n^b \hat T_{ab} = D_b K^b_a -D_a  K
\end{eqnarray}
%
in terms of $(q_{ab},n^a)$. At $i_0$ it becomes  
%
\begin{eqnarray}
0 \hat = D_b K^b_a - D_a K. \label{Co}
\end{eqnarray}
%
Substituting Eq. (\ref{K}) into Eq. (\ref{Co}), we see that 
%
\begin{eqnarray}
 D_a F \hat = 0
\end{eqnarray}
%
and then 
%
\begin{eqnarray}
F \hat = {\rm const.}
\end{eqnarray}
%
Since we can set $F\hat = 1$ without loss of generality, 
%
\begin{eqnarray}
K_{ab} \hat = q_{ab}. \label{K2}
\end{eqnarray}
%
Here, we used the gauge freedom of the conformal factor $\Omega\rightarrow\omega\Omega$, that is, since under this transformation $F$ transforms as 

\begin{eqnarray}
F\rightarrow F'\hat=\omega^{-2}F,
\end{eqnarray}
we may choose $\omega$ to satisfy $\omega=F^{\frac{1}{2}}$.
From the Gauss equation 
%
\begin{eqnarray}
\Omega^{-2} \hat e^a_I \hat e^b_J {}^{(n)}R_{ab}
 & = & \Biggl[ {}^{(n-1)} R_{ab}-K  K_{ab}
-F^{\frac{1}{2}}K_{ab} \nonumber \\
& & + 2 K_{ac}  K^c_b-F^{\frac{1}{2}} D_a  D_b F^{-\frac{1}{2}} \nonumber \\
& & +\Omega F^{-\frac{1}{2}} \mbox \pounds_{n} K_{ab} 
\Biggr]  e^a_I  e^b_J, 
\end{eqnarray}
%
we have 
%
\begin{eqnarray}
{}^{(n-1)} R_{ab} \hat = (n-2)  q_{ab}
\end{eqnarray}
%
and then 
%
\begin{eqnarray}
{}^{(n-1)} R_{abcd} \hat = {}^{(n-1)} C_{abcd}+2  q_{a[c}  q_{d]b}. \label{main}
\end{eqnarray}
%
This is simple but the main consequence in our paper. In four dimensions, 
due to the absence of the three-dimensional Weyl tensor ${}^{(3)}C_{abcd}=0$, 
%
\begin{eqnarray}
{}^{(3)} R_{abcd} \hat = 2  q_{a[c}  q_{d]b}.
\end{eqnarray}
%
This implies that $i_0$ is a three-dimensional unit hyperboloid. In the case of 
$n \geq 5$, the situation is drastically changed because ${}^{(n-1)}C_{abcd} \neq 0$ 
in general. Indeed, we have an $n$-dimensional solution with non-zero Weyl tensor 
as shown in the next section. Such spacetimes are not included in the category of 
asymptotically flat spacetimes.

\section{Static spacetimes}

In this section, we focus on static spacetimes in higher dimensions. To investigate 
the asymptotic structure, it is better to adopt a definition separately. 

In the static spacetime, the metric can be written as 
%
\begin{eqnarray}
ds^2=-V^2 dt^2 +q_{ij}dx^i dx^j
\end{eqnarray}
%
where $i,j=1,2, \cdots, n-1$. The Einstein equation becomes 
%
\begin{eqnarray}
{}^{(n)}\hat R_{\hat 0 \hat 0} = \frac{1}{V}D^2V=\hat T_{\hat 0 \hat 0}+\frac{1}{n-2} \hat T
\end{eqnarray}
%
and
%
\begin{eqnarray}
{}^{(n)}\hat R_{ij} = {}^{(n-1)}R_{ij}-\frac{1}{V}\hat D_i \hat D_j V =
\hat T_{ij}-\frac{1}{n-2}g_{ij} \hat T.
\end{eqnarray}
%

\subsection{Structure of spatial infinity in static slices}

{\it Definition.}
Physical static slice $(\hat \Sigma, \hat q_{ab})$ has a spatial infinity $\tilde i_0$ if there is a 
smooth function $\Omega$ satisfying the following features (i), (ii) and an appropriate 
fall off condition for the energy-momentum tensor.

(i) $\Omega \hat =0$ and $ d \Omega| \hat{\neq} 0$ 

(ii)The following quantities have smooth limits on $\tilde{i}_0$:
%
\begin{eqnarray}
 h_{ab}=\Omega^2 (\hat q_{ab}-\Omega^{-4}F^{-1}\hat \nabla_a \Omega \hat \nabla_b \Omega)
=\Omega^2 \hat h_{ab} 
\end{eqnarray}
%
%
\begin{eqnarray}
n^a: = \Omega^{-4} \hat g^{ab} \hat \nabla_b \Omega,
\end{eqnarray}
%
where
%
\begin{eqnarray}
F= \Omega^{-4}\hat g^{ab} \hat \nabla_a \Omega \hat \nabla_b \Omega = \mbox \pounds_n \Omega. 
\end{eqnarray}
%
$h_{ab}$ has the signature $(+,+,\cdots,+)$. 

The extrinsic curvature defined by 
%
\begin{eqnarray}
\hat k_{ab} = \frac{1}{2} \mbox \pounds_{\hat n} \hat h_{ab} 
\end{eqnarray}
%
is singular at $\Omega=0$. In the same way as the previous section, we 
define $k_{ab}= \Omega \hat k_{ab}$ and then we see that $k_{ab} \hat = h_{ab}$ 
from the Codacci equation. From the Gauss equation, ${}^{(n-2)}R_{ab}\hat 
=(n-3)h_{ab}$.  Thus
%
\begin{eqnarray}
{}^{(n-2)}R_{abcd}\hat = {}^{(n-2)}C_{abcd}+2 h_{a[c}h_{d]b}. \label{main2}
\end{eqnarray}
%

In five or four dimensional spacetimes, 
%
\begin{eqnarray}
{}^{(3,2)}R_{abcd}\hat = 2 h_{a[c}h_{d]b}
\end{eqnarray}
%
It represents a three or two-sphere. 

\subsection{Multipole moments}

In this subsection we define the multipole moments in a covariant way. To do so it is 
better to change the formalism and use the conformal completion defined by Geroch \cite{Geroch}. 

{\it Definition.} A 
physical static slice $(\hat \Sigma, \hat q_{ab})$ has a spatial infinity $\tilde i_0$ if there is a 
smooth function $\Omega$ such that
%
\begin{eqnarray}
\Omega \hat = 0, ~~\tilde \nabla_a \Omega \hat = 0~~{\rm and}~~\tilde \nabla_a \tilde \nabla_b \Omega \hat{\neq} 0
\end{eqnarray}
%
and
%
\begin{eqnarray}
\tilde q_{ab}= \Omega^2 \hat q_{ab}. 
\end{eqnarray}
%
has a smooth limit on $\tilde i_0$. 

As an example, consider Euclid space. The metric is 
%
\begin{eqnarray}
d\ell^2 = dr^2+r^2 d\Omega_{n-2}
\end{eqnarray}
%
$\Omega$ is taken to be $\Omega = r^{-2}$. Then  
%
\begin{eqnarray}
\Omega^2 d\ell^2 = r^{-4} dr^2+r^{-2} d\Omega_{n-2} = dR^2+R^2 d\Omega_{n-2}
\end{eqnarray}
%
where $R=r^{-1}$. Then $\tilde i_0$ is just the center in an unphysical slice. 
Moreover, $ \tilde \nabla_a \Omega = 2R \tilde \nabla_a R \hat = 0$ and  
$ \tilde \nabla_a \tilde \nabla_b \Omega \hat =2 \tilde \nabla_a R \tilde \nabla_b R \hat{\neq} 0$. 

Following Geroch argument, we might be able to identify the values of the following tensor at spatial infinity as 
multipole moments 
\begin{widetext}
%
\begin{eqnarray}
& & P = \frac{1}{2} (1-V)\Omega^{-\frac{n-3}{2}} \nonumber \\
& & P_{a_1 a_2 \cdots a_{s+1}}  =  {\cal O}\Biggl[ \tilde \nabla_{a_1} P_{a_2 a_3 \cdots a_{s+1}} 
-\frac{s(2s+n-5)}{2(n-3)}  {}^{(n-1)}\tilde R_{a_1 a_2} P_{a_3 a_4 \cdots a_{s+1}} \Biggr],  \label{moment}
\end{eqnarray}
%
where ${\cal O}[T_{a_1 a_2 \cdots a_r}]$ denotes the totally symmetric, trace free parts of 
$T_{a_1 a_2  \cdots a_r}$. This is recursive and a coordinate-free definition. 
The definition relies on the argument of the conformal rescaling($\Omega' = \Omega \omega$) \cite{Geroch} 
( The multipole moments in Newtonian system depend on the choice of the origin of the coordinate. 
This behavior of the multipole moments is reflected by the transformation of the multipole moments 
under a change of the conformal factor. The second term in the above definition reflects this in 
curved spacetimes.) Since the rescaling corresponds to a translational transformation, we 
wish the following transformation for $P_{a_1 a_2 \cdots a_{s+1}}$
%
\begin{eqnarray}
P'_{a_1 a_2 \cdots a_{s+1}}  =  P_{a_1 a_2 \cdots a_{s+1}} 
 -\frac{(2s+n-3)(s+1)}{2} {\cal O} \Biggl[ P_{a_1  \cdots a_s } \tilde \nabla_{a_{s+1}} \omega \Biggr]. 
\label{trans}
\end{eqnarray}
%
\end{widetext}
We can check that it indeed holds for the definition of Eq. (\ref{moment}). Note that 
the definition dose not contain the Weyl tensor ${}^{(n-1)}\tilde C_{abcd}$. 

In four-dimensional asymptotically flat spacetimes, we can show that they become identical with 
the coordinate dependent 
multipole moments defined by Thorne \cite{Kip,Gursel}. And most important feature is that 
stationary and vacuum spacetimes 
having the same multipole moments are isometric with each other in four dimensions. That is, the local 
structure of the stationary and vacuum 
spacetimes is completely 
determined by the multipole moments \cite{Beig}. In Newtonian gravity, this fact is trivial. However, in general relativity, 
it is not so. As demonstrated by an example below, the situation will be drastically changed in higher dimensions. 

There are generalized Schwarzschild spacetimes and the metric is \footnote{The form is similar to that in 
the isotropic coordinates of the Schwarzschild solution. If we use the coordinate $\rho = r h(r)^{\frac{2}{n-3}}$, 
the metric becomes the familiar form: $ds^2= -F(\rho)dt^2+F(\rho)^{-1}d\rho^2+\rho^2 \sigma_{AB}dx^A dx^B$, where 
$F(\rho)= 1-4(\mu/\rho)^{n-3}$. }
%
\begin{eqnarray}
ds^2=-f(r)^2 dt^2+ h(r)^{\frac{4}{n-3}}\Biggr[ dr^2+r^2 \sigma_{AB}dx^A dx^B \Biggr]. 
\end{eqnarray}
%
where $A,B=2,3,\cdots, n-1$. $f(r)$ and $h(r)$ are given by 
%
\begin{eqnarray}
f(r)=\frac{1-(\frac{\mu}{r})^{n-3}}{1+ ( \frac{\mu}{r})^{n-3}}.
\end{eqnarray}
%
and
%
\begin{eqnarray}
h(r)=1+ \biggl( \frac{\mu}{r} \biggr)^{n-3}.
\end{eqnarray}
%
$\sigma_{AB}$ is the metric of the Einstein space, that is, it obeys 
%
\begin{eqnarray}
{}^{(n-2)}R_{AB}(\sigma)=(n-3)\sigma_{AB},
\end{eqnarray}
%
where ${}^{(n-2)}R_{AB}(\sigma)$ is the Ricci tensor of $\sigma_{AB}$. 
The metric $\sigma_{AB}$ found by Bohm is given by \cite{Bohm}
%
\begin{eqnarray}
\sigma_{AB}dx^A dx^B = d \theta^2 + a^2 (\theta) d\Omega_{p} + b^2(\theta) d\Omega_{n-3-p},
\end{eqnarray}
%
where $5 \leq n-3 \leq 9$ with $p \geq 2$ and $q:=n-3-p \geq 2$. 
See Refs. \cite{Sean,Ishibashi} for the stability of such spacetimes.  

Taking $\Omega= ( \frac{1-V}{2} )^{\frac{2}{n-3}}$, 
the unphysical metric $\tilde q$ becomes 
%
\begin{eqnarray}
\tilde q = \Omega^2 \hat q = 
\Biggl(\frac{\mu}{r} \Biggr)^4 \Biggl[dr^2+ r^2 \sigma_{AB}dx^A dx^B \Biggr]. 
\end{eqnarray}
%
Defining
%
\begin{eqnarray}
R:=\mu^2 r^{-1}
\end{eqnarray}
%
then 
%
\begin{eqnarray}
\tilde q  = dR^2 + R^2 \sigma_{AB}dx^A dx^B. 
\end{eqnarray}
%
For this metric, the Ricci tensor becomes
%
\begin{eqnarray}
{}^{(n-1)} \tilde R_{ij}=0,
\end{eqnarray}
%
Finally, we can see at spatial infinity, 
%
\begin{eqnarray}
& & P =\frac{1}{2}(1-V)\Omega^{-\frac{n-3}{2}}=1 \nonumber \\
& & P_{a_1 a_2 \cdots a_s}=0~~{\rm for}~~ s \geq 1. 
\end{eqnarray}
%
Thus this spacetime has the same multipole moments as spherical Schwarzschild spacetimes. We cannot 
distinguish them from one another using only multipole moments. This problem comes from the absence of the 
Weyl tensor in the definition Eq. (\ref{moment}). Because of the total anti-symmetricity 
of the Weyl tensor, there is no room for the Weyl tensor in the definition. We need the information related to the Weyl tensor independently. 
Hence, we might be able to 
expect that we can uniquely specify higher dimensional spacetimes by the multipole moments and Weyl tensor. 
 The Bohm metric has the 
following non-trivial Weyl tensor ${}^{(n-2)}C_{abcd}$ \cite{Sean}:
%
\begin{eqnarray}
& & {}^{(n-2)}C_{\hat \theta \hat A_1 \hat \theta \hat A_2}=c_1(\theta)  \delta_{\hat A_1 \hat A_2}, \nonumber \\
& & {}^{(n-2)}C_{\hat \theta \hat B_1 \hat \theta \hat B_2}= c_2(\theta) \delta_{\hat B_1 \hat B_2} \nonumber \\
& & {}^{(n-2)}C_{\hat A_1 \hat B_1 \hat A_1 \hat B_2}= c_3 (\theta) \delta_{\hat A_1 \hat A_2} 
\delta_{\hat B_1 \hat B_2}, \\
& & {}^{(n-2)}C_{\hat A_1 \hat A_2 \hat A_3 \hat A_4}=2 c_4 (\theta) \delta_{\hat A_1 [ \hat A_3} 
\delta_{\hat A_4 ] \hat A_2}, \nonumber \\
& & {}^{(n-2)}C_{\hat B_1 \hat B_2 \hat B_3 \hat B_4}= 2 c_5(\theta) \delta_{\hat B_1 [ \hat B_3} 
\delta_{\hat B_4 ] \hat B_2}, \nonumber 
\end{eqnarray}
%
where 
%
\begin{eqnarray}
& & c_1 (\theta) = -1-\frac{a''}{a},~~c_2(\theta)= -1-\frac{b''}{b}, \nonumber \\
& & c_3 (\theta) = -1-\frac{a'}{a} \frac{b'}{b}, \\
& & c_4 (\theta) = \frac{1-a'^2-a^2}{a^2},~~c_5 (\theta)= \frac{1-b'^2-b^2}{b^2}. \nonumber 
\end{eqnarray}
%
In the above we used the orthogonal basis $\lbrace \hat e^{A_1}, \hat e^{A_2}, \cdots \hat e^{A_p} \rbrace $ 
and $\lbrace \hat e^{B_1}, \hat e^{B_2}, \cdots \hat e^{B_{n-3-p}} \rbrace $, that is, 
%
\begin{eqnarray}
& & a^2(\theta) d\Omega_p = \delta_{A_1 A_2} \hat e^{A_1} \otimes \hat e^{A_2}, \nonumber \\
& & b^2(\theta) d\Omega_{n-3-p} = \delta_{B_1 B_2} \hat e^{B_1} \otimes \hat e^{B_2}, \nonumber 
\end{eqnarray}
%
where $A_1,A_2, \cdots= 3, 4, \cdots, p+2$ and $B_1, B_2, \cdots = p+3, p+4, \cdots, n-1$.

\section{Summary and Discussion}

In this paper, we investigated the asymptotic structure at spatial infinity in 
higher dimensional spacetimes. One will realize that this is quite important when one tries 
to perform numerical computations or prove the uniqueness theorem. This is because one must impose 
asymptotic boundary conditions on them. In higher dimensions, it turned out that there are 
many varieties. That is, it is unlikely that the asymptotic symmetry is raised automatically due to 
the non-trivial Weyl tensor(See Eqs. (\ref{main}) and (\ref{main2})) at spatial infinity. 
If and only if we set the Weyl tensor to zero, the asymptotic 
flatness seems to be guaranteed. Since the definition of the multipole moments cannot 
include the Weyl tensor part, the static solutions are degenerate in terms of multipole moments. 
We must at least use the Weyl tensor if one wants to split these solutions. This is contrasted 
with the four dimensional spacetimes where the local structure of static and vacuum solutions can be 
uniquely figured out from the higher moments. The point is just the dimension. From our study we must 
specify the multipole moments $P_{a_1 a_2 \cdots a_r}$ and Weyl tensor ${}^{(n-2)}C_{abcd}$ 
for each individual solution. This is a lesson for the boundary condition in a 
numerical study and gives us an insight into the argument of the uniqueness theorem. Therein we 
should carefully think of the Weyl tensor or something similar.  

Now, we might be able to have the following conjecture: 
\vskip 2mm
{\it If the two static vacuum spacetimes defined here have the same multipoles and 
Weyl tensor ${}^{(n-2)}C_{abcd}$ at spatial infinity, they are isometric in a local sense.}  
\vskip 2mm
There are many remaining issues; First of all, the details of the structure of spatial infinity. 
It is unlikely that there is asymptotic symmetry because of the lack of maximal symmetry. 
Even if this is so, it is important to ask why asymptotic symmetry cannot exist. The next issue 
is the proof of 
the statement that the static spacetimes can be uniquely specified by higher multipole 
moments and the Weyl tensor. We can also extend our argument on static spacetimes to 
stationary cases. Since the Weyl tensor appears in our conjecture, the relation with the peeling 
theorem associated with null infinity is also interesting (See Ref. \cite{peel} for the peeling 
theorem in four dimensions.).
Finally, in four dimensions, since the Weyl curvature on the static slices vanishes, it, of course, never 
contributes to the multipole moments. However, in more than four dimensions, the Weyl curvature on spacelike 
hypersurfaces in general does not vanish. Since the multipole moments imply deviation from spherical 
symmetry, they  seem to contain the Weyl tensor. 
We may be able to extend Geroch's definition of multipole moments (\ref{moment}) to a refined 
form to contain the Weyl curvature in higher dimensional space-time.

\section*{Acknowledgments}
We would like to thank D. Ida, A. Ishibashi, 
K. Nakao, M. Sasaki and M. Siino for useful discussion and comments. 
The work of TS was supported by Grants-in-Aid for Scientific
Research from the Ministry of Education, Science, Sports and Culture of 
Japan(No.13135208, No.14740155 and No.14102004). The work of ST was supported 
by the 21st Century COE Program at TokyoTech "Nanometer-Scale Quantum Physics" supported 
by the Ministry of Education, Culture, Sports, Science and Technology.



\end{document}